\documentclass{sigchi}

\toappear{\small Permission to make digital or hard copies of all or part of this work for personal or classroom use is granted without fee provided that copies are not made or distributed for profit or commercial advantage and that copies bear this notice and the full citation on the first page. To copy otherwise, or republish, to post on servers or to redistribute to lists, requires prior specific permission and/or a fee.
\newline 
\emph{WebSci’14}, June 24 – June 26, 2014, Bloomington, Indiana. \\
ACM 978-1-4503-1889-1.
}
\usepackage{balance}  
\usepackage{graphics} 
\usepackage{times}    
\usepackage{url}      

\makeatletter
\def\url@leostyle{%
  \@ifundefined{selectfont}{\def\UrlFont{\sf}}{\def\UrlFont{\small\bf\ttfamily}}}
\makeatother
\def\pprw{8.5in}
\def\pprh{11in}

\usepackage[pdftex]{hyperref}
\hypersetup{
pdftitle={WebScience Conference Proceedings Format},
pdfauthor={LaTeX},
pdfkeywords={WebScience, proceedings, archival format},
bookmarksnumbered,
pdfstartview={FitH},
colorlinks,
citecolor=black,
filecolor=black,
linkcolor=black,
urlcolor=black,
breaklinks=true,
}

\usepackage{graphicx}

\makeatletter
\newif\if@restonecol
\makeatother

\usepackage[vlined, ruled, commentsnumbered, linesnumbered]{algorithm2e}
\SetKwInput{KwInit}{Initialize}
\SetKwInput{KwInput}{Input}
\SetKwInput{KwOutput}{Output}
\SetKwFor{parfor}{parfor}{:}{end parfor}

\begin{document}
%
\conferenceinfo{GeoRich}{2014 Salt Lake City, Utah USA}
\CopyrightYear{2014} 

\title{Inferring the geographic focus of online documents from social media sharing patterns}

\numberofauthors{2} 
%
\author{
\alignauthor
Ryan Compton\\
       \affaddr{HRL Laboratories}\\
       \email{rfcompton@hrl.com}
\alignauthor
Matthew S. Keegan\\
       \affaddr{HRL Laboratories}\\
       \email{mskeegan@hrl.com}
\alignauthor
Jiejun Xu\\
       \affaddr{HRL Laboratories}\\
       \email{jxu@hrl.com}
}

\maketitle
\begin{abstract}
Determining the geographic focus of digital media is an essential first step for modern geographic information retrieval. However, publicly-visible location annotations are remarkably sparse in online data. In this work, we demonstrate a method which infers the geographic focus of an online document by examining the locations of Twitter users who share links to the document.

We apply our geotagging technique to multiple datasets built from different content: manually-annotated news articles, GDELT, YouTube, Flickr, Twitter, and Tumblr. 

\end{abstract}

\category{H.3}{Information storage and retrieval}{Indexing methods}


\keywords{Data mining, Information retrieval, Geolocation}

\section{Introduction}
\label{sec:introduction}

Despite its importance, geographic information is only sparsely available in public online digital media. To facilitate geographic information retrieval tasks, content-specific geotagging methods have been developed by researchers from a wide variety of fields, ranging from natural language processing \cite{Leidner2004}, to computer vision \cite{serdyukov2009placing}. In this work, we investigate the possibility of content-agnostic geotagging; i.e. geotagging from the url alone.

Our aim is to quantify how effectively one can infer the geographic focus of an online document by examining the locations of Twitter users who share links to the document. This will open up the possibility of high-volume digital media geotagging which can be accomplished without content-specific expertise.

Given a url, and a high-coverage database of Twitter user locations, we assign a location to the url using the median location of the Twitter users who have shared the url. To be specific, let $\mathcal{G}$ be the set of locations of Twitter users who share the url and $d(\cdot , \cdot)$ geodesic distance measured via Vincenty's formulae, our geotag is the $l1$-multivariate median \cite{Vardi2000} of the locations in $\mathcal{G}$:
\begin{subequations} \label{eq:l1median} 
\begin{align}
\text{geotag} &= \text{median} (\mathcal{G})  \\
& =  \underset{x}{\operatorname{\text{argmin}}} \; \sum_{y \in \mathcal{G}}  d(x,y) 
\end{align}
\end{subequations}

Online documents are not necessarily geographically focused. In order to identify documents for which geotagging is appropriate, we use a robust estimate of the dispersion of locations in $\mathcal{G}$ and do not attempt to geotag documents for which the dispersion is beyond a user-specified threshold. We quantify uncertainty in our geotags using the median absolute deviation of the locations in $\mathcal{G}$:
\begin{subequations} \label{eq:l1spread}
\begin{align}
\text{uncertainty} &=\text{dispersion}(\mathcal{G}) \\
& = \underset{x \in \mathcal{G}}{\operatorname{\text{median}}}\left( d(x, \text{median}(\mathcal{G})) \right)
\end{align}
\end{subequations}

Due to the high number of outliers present in geographic social media, use of robust estimates for the center and spread of $\mathcal{G}$ is essential for our work. Indeed, research into Twitter geographic patterns based on non-robust statistics has concluded that there is little agreement between user activity and geography \cite{leetaru2013mapping}. On the other hand, research involving robust statistics  has discovered strong geographic underpinnings \cite{jurgens2013s} \cite{Yamaguchi:2013:LUL:2512938.2512941}.

For text geotagging, we evaluate our method on news documents pertaining to localized events as well as tweets containing mentions of unambiguous toponymns. We find that we are able to infer the geographic focus of news media with some accuracy. Our results on Twitter are, surprisingly, nearly an order of magnitude better. Considering that natural language processing tools are well-developed for news but are only recently being adapted to  social media, we suspect that our ability correctly infer the geographic content of tweet will have many applications. For example, when an action and a future date are mentioned but no location can be easily discerned, the technique presented here allows one to infer the location of an upcoming event \cite{compton2013detecting}. 

Online photos and videos are rarely annotated with public location data. For example, of the $447,441$ YouTube links geotagged with our method, only $14,583$ $(3\%)$ were given a location by the YouTube data API \footnote{\url{https://developers.google.com/youtube/v3/}}. On Flickr we found approximately $14\%$ of photos could be geotagged with the Flickr API \footnote{\url{https://www.flickr.com/services/api/}}.

\section{Background and Related Work}
\label{sec:background}
By far the most common technique to geotag digital media \cite{Kelm2013} is through a process of geo-parsing text, identifying toponyms in text or metadata, and toponym resolution, mapping from a name of a place to an unambiguous geographic location \cite{Leidner2004}. With respect to digital media, geotagging using text has been applied to general webpages \cite{McCurley2001,Amitay2004}, Wikipedia pages \cite{Overell2008} and visual media such as photos posted on Flickr with geo-referenced tags \cite{Ahern2007}. 

Toponyms are usually stored in a dictionary called a gazetteer. Toponym resolution is a challenging task for a number of reasons: misspellings; the use of abbreviation, which is particularly prevalent in social media microtext services such as Twitter; evolving new and changing location names, for example through urban construction or expansion or through changing vernacular; and, through ambiguities in the use of toponyms. Geocoding literature focuses particularly on the case of topoonym ambiguities of toponyms, of which there are two types \cite{Amitay2004}: geo/geo, where a given toponym can be used to describe multiple distinct locations, and geo/non-geo, where a toponym also has a non-geographic meaning.

Natural language processing and machine learning techniques are used to identify mispellings and abbreviations and to add local terms for automatic gazetteer enrichment \cite{Lieberman2010,Gelernter2013,Gelernter2013a}. Methods for disambiguation include modeling co-occurrences of toponyms on Wikipedia \cite{Overell2008}.

Other approaches construct their own generative, realtional location dictionaries by data-mining rich, online geo-referenced data, such as using geo-tags on Flickr \cite{Rattenbury2009, VanLaere2011}. The work of \cite{Crandall2009} extends these by adding temporal and image features to the textual cues to geotag images using a multi-feature approach. The work of \cite{VanLaere2014} and \cite{OHare2013} improve these approaches by defining location priors and, in the case of \cite{VanLaere2014}, using statistical term selection. The joint modelling of location-based and topic-based content on social media posts in \cite{Ahmed2013} is used to build geographic language models, that give better geotagging results than geographic terms alone, and in \cite{Yin2011} to group geographic regions by topic-similarity.

The work in \cite{Ireson2010} uses co-occurrences of toponyms in tags of geotagged Flickr photos, but expands the scope of closeness between toponyms to include tags made by the same user and by contacts within the user's social network. This assumes that links in a social network strongly correlate to users' location, in other words that closeness centrality in the social network is significant. This is not an assumption that we have seen in other toponym resolution publications, but it is a central thesis of our proposed methods. The results of \cite{Ireson2010} show that toponym resolution is improved by including the posts of contacts, but that adding the posts of contacts of contacts does not perform as well.

\section{Datasets}
\label{sec:datasets}

A key ingredient of our approach is an accurate and high-coverage Twitter user location database. To build this, we employ a geotagging algorithm which iteratively infers the location of a Twitter user by examining the locations of their online friends \cite{compton2014geotagging} \cite{jurgens2013s}.

\subsection{Twitter user geolocation}
\label{sec:usergeo}

Physical locations of Twitter users are only sparsely available in public data. Less than 1\% of tweets are annotated with GPS and less than 30\% of users report unambiguous locations in their profiles \cite{leetaru2013mapping}.  

As a result, several Twitter geotagging methods have been developed in recent years; many based on natural language processing \cite{mahmud2012tweet}. The distinguishing feature of the method outlined here is that it is based solely on social network analysis and is thus language-agnostic and straightforward to scale.

An appropriate social network is a fundamental part of the algorithm. Twitter users often ``@mention'' each other by appending an ``@'' to the mentioned user's name. We build an undirected social network, $G=(V,E)$, with users as vertices and reciprocated @mentions between users as edges. The key advantage to constructing a social graph from @mentions (as opposed from ``followers") is that it enables us to build a large social graph from a large collection of publicly-visible tweets without being burdened by Twitter API rate limiting.

We use a $10\%$ sample of public tweets collected between April 2012 and January 2014\footnote{\url{http://gnip.com/}}. This amounts to $67.2$TB of json data (uncompressed) and $22,455,584,506$ @mentions. From the complete set of @mentions we built a weighted and directed network of $7,266,222,329$ edges by condensing multiple mentions into weighted edges. Filtering down to only reciprocated edges leaves us with a network of $915,189,977$ edges\footnote{i.e. a $12.6\%$ chance of @mention reciprocation} and $101,230,940$ users. This network is the focus of our experiments.

We define a function, $f$, which assigns to each user an estimate of their physical location. Users may opt to make their location publicly available though cellphone-enabled GPS or self-reported profile information. For this small set of users, computation of $f$ is relatively straightforward: to assign a unique location to a user from the set of their GPS-tagged tweets, we compute the $l1$-multivariate median (cf. (\ref{eq:l1median})) of the locations they have tweeted from.

Additionally, we extract self-reported home locations by searching through a list obtained via \url{www.geonames.org} for exact matches in user profiles. When self-reporting users also reveal their location though GPS, we opt to use their GPS-known location. We remove self-reports which have not appeared in over one month. This provides us with home locations for $9,466,251$ users. The median discrepancy between our self-reported location extraction and GPS is $7.69$ km.

We regard users who reveal their location via GPS or self-reports as our ground truth. The total number of such users is $17,589,170$. Denote this set of users by $L$, and the remaining users in the network by $U$. The vertex set is thus partitioned as
\begin{equation}
V = L + U
\end{equation}
and our goal is to assign a value of $f$ to nodes in $U$.

Work done in \cite{jurgens2013s} and \cite{compton2014geotagging} demonstrates that Twitter @mentions predominately occur over short geographic distances. Ties in an ideal social network, therefore, will span the smallest possible geographic distance. Thus to infer locations for the unlabelled users, $U$, \textbf{we seek a network such that the sum over all geographic distances between connected users is as small as possible}. This is accomplished by solving the optimization:
\begin{subequations}
\begin{alignat}{2}
\label{eq:tvopt}
\min_{\mathbf{f}} & | \nabla \mathbf{f} | \\
\text{subject to: } & f_i = l_i \; \text{for} \; i \in L,\\
& \text{and} \; \text{dispersion}(\nabla f_i) \leq \gamma
\end{alignat}
\end{subequations}
where $\mathbf{f} = (f_1,\ldots,f_N)$ is a vector encoding location estimates for each user, $\text{dispersion}(\nabla f_i)$ is the median absolute deviation (cf. \ref{eq:l1spread}) of the users distances to their friends, the parameter $\gamma$ defines how dispersed we allow a user's friends to be, and the anisotropic total variation, $| \nabla \mathbf{f} |$, on the Twitter @mention network is defined by:
\begin{equation}
\label{tvdef}
| \nabla \mathbf{f} | = \sum_{ij} w_{ij} d(f_i,f_j)
\end{equation}
The edge weights, $w_{ij}$, are equal to the minimal number of reciprocated @mentions between users $i$ and $j$. We set the parameter $\gamma$ to 100km in our code.

This Twitter user geotagging technique falls under the category of transductive learning and shares some similarity with ``label propagation'' \cite{zhu2002learning}. However, unlike label propagation, our labels (latitude/longitude pairs) are continuously valued. Equation (\ref{eq:tvopt}) exploits this additional structure by taking into account geodesic distance. Total variation has demonstrated remarkable performance as an optimization heuristic for several information inference problems across a wide variety of fields \cite{Rudin1992} \cite{Smith:2010:IDE:1840693.1928520} \cite{bresson2013multiclass}.

\begin{algorithm}
\KwInit{$f_{i} = l_{i}$ for $i\in L$ and parameter $\gamma$}
\For{$k = 1 \ldots N$}{
\parfor{$i$}{
\eIf{$i \in L$}{
$f^{k+1}_i = l_i$
}{
	\eIf{$\overset{\sim}{\nabla f_i} \leq \gamma$}
		{$f^{k+1}_i = \underset{f}{\operatorname{\text{argmin}}} |\nabla_i(\mathbf{f}^k,f)|$}
		{no update on $f_i$}
}
}
$\mathbf{f}^{k} = \mathbf{f}^{k+1}$
}
\caption{Parallel coordinate descent for dispersion-constrained total variation minimization.}
\label{alg:pcd2}
\end{algorithm}

The total variation functional is nondifferentiable. Solving a total variation-based optimization is thus a formidable challenge and vastly different methods have been proposed for several decades \cite{Goldstein}. We employ ``parallel coordinate descent'' to solve (\ref{eq:tvopt}). Most variants of coordinate descent cycle through the domain sequentially, updating each variable and communicating back the result before the next variable can update. The scale of the data we work with necessitates a parallel approach, prohibiting us from making all the communication steps required by a traditional coordinate descent method.

At each iteration, our algorithm simultaneously updates each user's location with the $l1$-multivariate median of their friend's locations. Only after all updates are complete do we communicate our results over the network. Note that the argument that minimizes $|\nabla_i(\mathbf{f}^k,f)|$ is the $l1$-multivariate median of the locations of the neighbours of node $i$. In other words, we iteratively update each user's location with the median of their friends locations, provided that their friends are not too dispersed.

Empirically, alg. \ref{alg:pcd2} converges, providing us with estimates of locations for $\mathbf{91,984,163}$ Twitter users. Leave-many-out cross validation with a $10\%$ hold-out set from the GPS-known users yields a median error of \textbf{6.65km}. It should be noted that outlying errors are indeed present; the mean error is 300.06km and the standard deviation is 1,131.83km.

\subsection{Toponyms in social media}
\label{sec:tweet_content_geo}

The goal of the present work is to infer the geographic focus of an online document. The results of the previous section provide us with the geographic origin of a tweet, which is not necessarily the same.

To understand the geographic focus of a tweet, one straightforward method is to search for unambiguous toponyms appearing in the tweet's text. To evaluate the discrepancy between this simple heuristic and the approach proposed by this work, we must first construct a dataset of unambiguous toponyms which are used in Twitter text.

To this end, we search the self-reported location field in user profiles of GPS-annotated tweets for exact matches of toponyms found on \url{www.geonames.org} and filter out toponyms when the GeoNames location is far from the GPS tags. A similar technique was used by \cite{serdyukov2009placing} to identify geographic content in Flickr tags.

We examine a collection of $4,332,656,836$ tweets collected between 2013-12-29 and 2014-03-07. We found $53,233,310$ ($1.2\%$) tweets with both GPS tags and self-reported location fields. We found $67,467$ distinct GeoNames locations appearing in the self-reported location fields. To identify which of these locations are unambiguous, we filter out locations when less than $5$ different users have listed the location in their profile or the median discrepancy between the GPS tags and the GeoNames location is greater than $50$km. Short GeoNames entries (such as ``LA'') are often ambiguous, we remove any toponym comprised of less than $5$ characters from our list. This leave us with a set $14,250$ toponyms which are unambiguous in Twitter.

\subsection{Twitter/Tumblr User Alignment and Geolocation}
\label{sec:useralign}

Direct links to Tumblr posts are rare in Twitter. However, when users possess both Twitter and Tumblr account we can align the two with a simple approach based on url extractions \cite{Xu_websci2014}. This allows us to build a database of Tumblr user locations. By examining reblogs rather than retweets we can geotag a Tumblr post via (\ref{eq:l1median}) and (\ref{eq:l1spread}).

To align Tumblr and Twitter accounts we scan $67.2$TB of Twitter data collected between April 2012 and January 2014. Each json contains the content of the tweet as well as the profile information of the corresponding Twitter user. Two types of user account mentions are searched in our work. The first type is the \textit{Explicit Self-reported} accounts. 
We search for every user profiles in our Twitter corpus with the regular expressions, which indicate different social media platforms. For instance, we searched for the pattern \url{http://[www.]*[a-zA-Z0-9-_].tumblr.com} for mentions of Tumblr user accounts. 
The second type of account mention is obtained through \textit{Implicit Cross-Links}. 
Many existing social media sites support content synchronization in order to reduce end user effort. This allows a user to submit a post from one platform, and publish the same content to all other social media accounts under him or herself. 
For the case of Twitter, there is usually a URL appended at the end of a tweet to indicate its origin. For example, if a post was original published in Tumblr, and the content was synchronized to Twitter. 
There would be a short URL encoded at the end of the tweet referring back to the Tumblr post (e.g., \url{http://tmblr.co/ZVxw1y15H\_Go3}).

\subsection{Geographically annotated news media}
\label{sec:news}
Both manual and automated annotations of news media are used in our evaluation. The topic of geographic focus is most relevant when considering articles pertaining to a single localized event.

A massive collection of machine-annotated news documents pertaining can be obtained freely from the GDELT database \cite{leetaru2013gdelt}. Though limited to English language news outlets only, the GDELT database contains millions of annotations\footnote{the annotations use the CAMEO coding scheme \cite{leetaru2013gdelt}} of articles (along with referring urls) describing various actions worldwide. Geotagging is accomplished by identifying both a city-specific toponym as well as the associated nation within the same article. For example, an article mentioning ``Santiago'' would not be geotagged unless it also contains ``Chile''.

A large collection of manually-annotated news documents has been developed by MITRE for use in IARPA's OSI program \cite{ramakrishnan2014beating}. These articles are written in Spanish, English, and Portuguese and describe protests, strikes, and other forms of general civil unrest across ten nations in Latin America. The goal of the OSI program is to issue predictions of the events described by these articles prior their occurrence, a task which is possible to accomplish by correctly mining Twitter data \cite{compton2013detecting}. In this work, we will evaluate our geotagging methodology on both the manually-annotated dataset as well as GDELT.

\begin{figure}
\centering
\includegraphics[width=.5\textwidth]{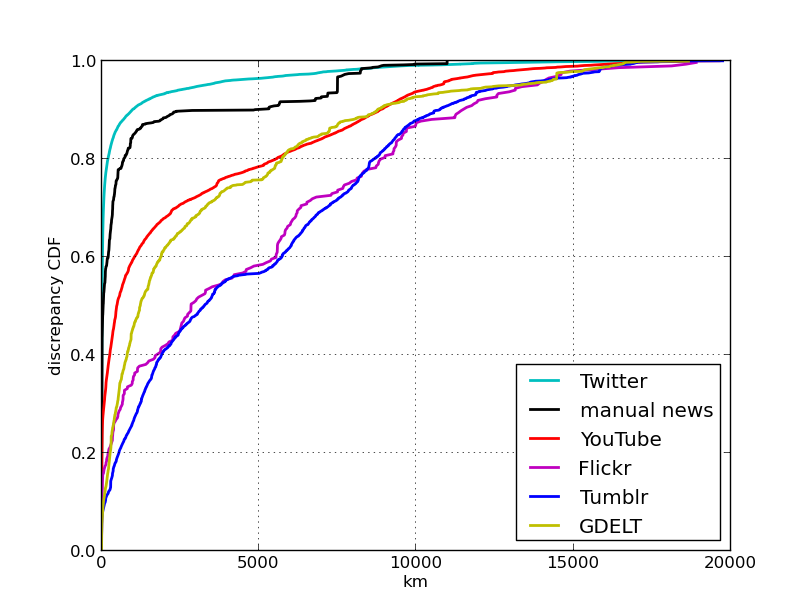}
\caption{Empirical CDF of discrepancies for the six different evaluation datasets with no restrictions on (\ref{eq:l1spread}). Here, our method is more accurate on Twitter and manually-annotated news data than on any other dataset. Note that after limiting (\ref{eq:l1spread}), accuracy on YouTube surpasses that of manually-annotated news (cf tbl. \ref{tbl:mederr_restricted}).}
\label{fig:all3_cdf}
\end{figure}

\section{Evaluation}
\label{sec:evaluation}

\begin{table}
\centering
\resizebox{\columnwidth}{!}{
    \begin{tabular}{|l|lll|}
    \hline
    & test points & median discrepancy & mean discrepancy \\ \hline
    Twitter & 12849 & 14.83 & 524.27 \\
    manual news & 1183 & 46.86 & 969.14 \\
    YouTube & 14582 & 483.20 & 2536.15 \\
    Flickr & 284 &  2840.04 & 4505.82 \\
    Tumblr & 7845 & 3256.78 & 4700.94 \\
    GDELT & 1580 & 1227.78 & 3039.39 \\ \hline
    \end{tabular}
    }
    \caption{Summary of discrepancy (in km), with no restrictions on (\ref{eq:l1spread}), between the proposed method and text or API-based geotagging on the six evaluation datasets.}
    \label{tbl:mederr}
\end{table}

\begin{table}
\centering
\resizebox{\columnwidth}{!}{
    \begin{tabular}{|l|lll|}
    \hline
    & test points & median discrepancy & mean discrepancy \\ \hline
    Twitter & 6864 & 6.70 & 110.04 \\
    manual news & 1115 & 36.66 & 902.01 \\
    YouTube & 5022 & 22.80 & 1001.58 \\
    Flickr & 42 &  371.88 & 2475.04 \\
    Tumblr & 1679 & 1371.21 & 3934.70 \\
    GDELT & 1580 & 304.74 & 2432.81 \\ \hline
    \end{tabular}
    }
    \caption{Summary of discrepancy (in km), with a $100$km restriction on (\ref{eq:l1spread}), between the proposed method and text or API-based geotagging on the six evaluation datasets. Imposing a restriction on (\ref{eq:l1spread}) improves accuracy and reduces coverage.}
    \label{tbl:mederr_restricted}
\end{table}

In fig. \ref{fig:all3_cdf} we plot the empirical cumulative distribution function of the discrepancy between the proposed approach and content-specific geotagging for all six datasets. A summary of statistics with no restrictions on (\ref{eq:l1spread}) is visible in tbl. \ref{tbl:mederr}; placing a $100$km restriction on (\ref{eq:l1spread}) leads produces the results in tbl. (\ref{tbl:mederr_restricted}). For all datasets, we require at least $3$ distinct geolocated users share the link before we assign a geotag. The effectiveness of our geotagging technique is highly domain specific.

\subsection{News media}

From the manually-annotated multilingual news corpus, we identified a set of $1,712$ urls which appeared a tweet. For GDELT, we were able to identify $2,328$ links in Twitter. For each of these urls, we scanned our 10\% Twitter datafeed for tweets containing the url and geotagged using (\ref{eq:l1median}).

While our median accuracy on the manually-annotated news media is acceptable for region-level inference, our ability to match the GDELT locations is substantially worse. One reason for the discrepancy with GDELT may be due to the fact that GDELT contains only English text, but describes events worldwide. This likely affects the location of users who are apt to share the articles. Alternatively, it could be due to the fact that the GDELT system is completely automated and has its own inaccuracies. We refer the reader to \cite{leetaru2013gdelt} for details on the geotagging methodology employed by GDELT.

Another concern is the presence of outliers. examining mean error indicates that outliers are a major concern. For all six datasets, the mean error reported in tbl.  \ref{tbl:mederr} is substantially higher than the median. As mentioned in the introduction, outliers are an unavoidable occurrence in this domain and need to kept in mind when working with geographic social media.

\subsection{Twitter content}

\begin{figure}
\centering
\includegraphics[width=.5\textwidth]{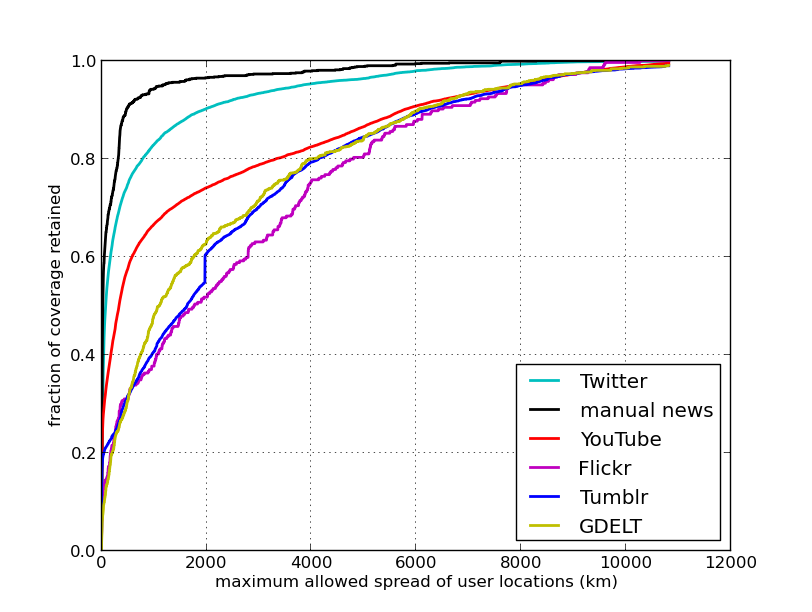}
\caption{Maximum allowed value of (\ref{eq:l1spread}) vs. coverage for the evaluation datasets. The manually-annotated news articles, Twitter, and YouTube all allow for high coverage geotagging even when restrictions on (\ref{eq:l1spread}) are tight.}
\label{fig:frac_retained}
\end{figure}

\begin{figure}
\centering
\includegraphics[width=.5\textwidth]{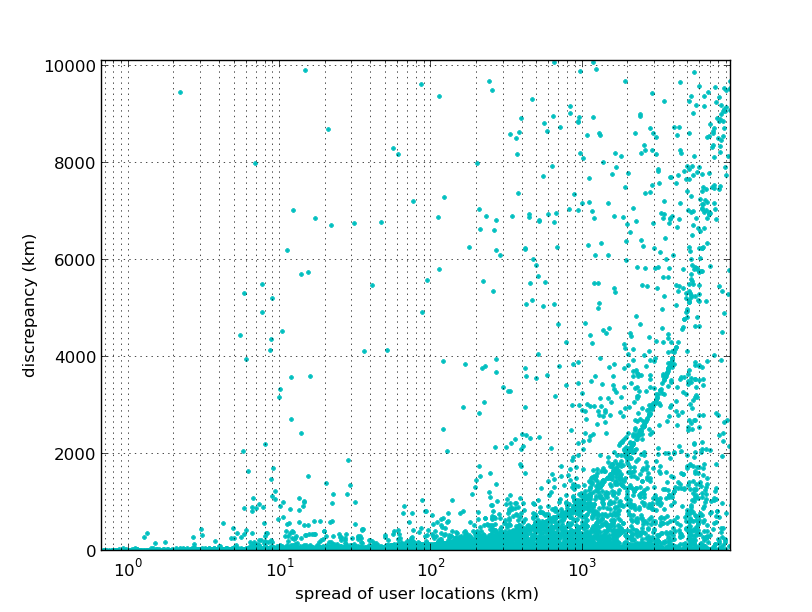}
\caption{Discrepancy vs. (\ref{eq:l1spread}) for the Twitter toponym data. The curve indicates that it is possible to improve accuracy by limiting the maximum tolerable dispersion. A  limitation below 100km will remove the majority of large discrepancies.}
\label{fig:dispersion_vs_error_log}
\end{figure}

We examine our $10\%$ Twitter feed for textual mentions of the $14,250$ toponyms described earlier. We only retain toponyms when at least $3$ different users have mentioned in tweet text, this leave us with $12,849$ distinct toponyms which we geotag using \ref{eq:l1median}.

Our ability to control geotagging error is highlighted in \ref{fig:dispersion_vs_error_log}, where we have plotted error vs dispersion for the $12,849$ toponyms. Note the log-scale, which is necessary to show the rarity of errors when restrictions on dispersion are tight. In \ref{fig:frac_retained} we study the coverage response as a function of $\gamma$. For both Twitter and manual news, coverage remains high until restrictions on $\gamma$ are under $1000$km.

\subsection{Tumblr content}

Tumblr posts are never annotated with publicly-visible GPS data. However, while most of the posts on Tumblr are images, several are text-based and contain mentions of the $14,250$ unambiguous toponyms discussed in the previous section. Here, for each post with mention of an unambiguous toponym, we report on the discrepancy between the mentioned location and the median of the reblogger locations. 

We examine $2,170,085,983$ (original and reblogs) Tumblr posts and find $133,451$ original posts with mention of an unambiguous toponym. Of these, $7,845$ were reblogged $3$ or more times by a Tumblr user in our Tumblr geolocation database. Using (\ref{eq:l1median}), we compute a social-network based geotag for these posts and report the discrepancy against the toponym-based geotag in tbl. \ref{tbl:mederr}. Restricting the maximum allowed dispersion of rebloggers via (\ref{eq:l1spread}) does little to reduce the discrepancy on Tumblr data (cf. tbl. \ref{tbl:mederr_restricted}). We conclude that, unlike @mentions in Twitter, reblogs on Tumblr are not grounded in geography.

\subsection{YouTube and Flickr data}
\label{sec:ytandfr}

Our method opens up the possibility of high-volume digital media geotagging which can be accomplished without content-specific expertise. In order to demonstrate the effectiveness of content-agnostic geotagging we apply our technique to urls referring to photos and videos.

\begin{figure}
\centering
\includegraphics[width=.5\textwidth]{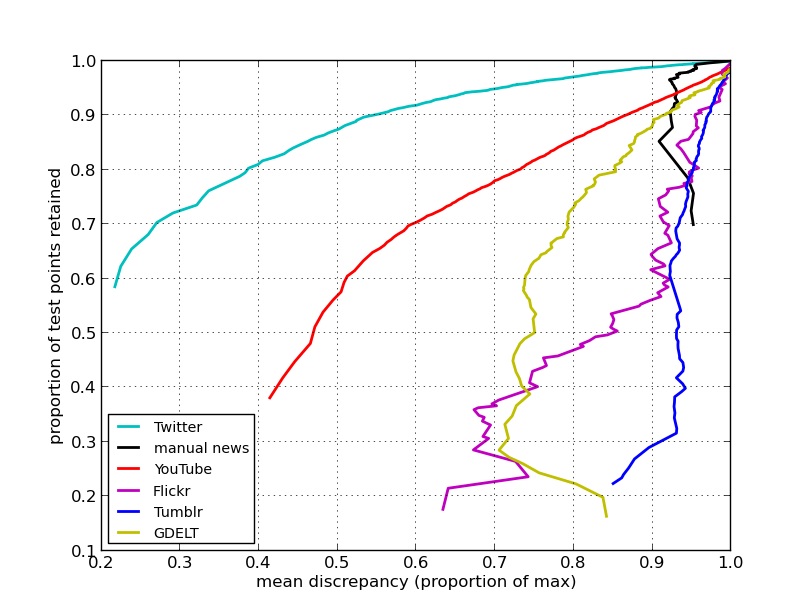}
\caption{Mean error characteristic curves for varying restrictions on (\ref{eq:l1spread}). For each restriction on (\ref{eq:l1spread}), we plot the mean error  obtained on the $x$-axis versus the proportion of points predicted on the $y$-axis. For example, we can reduce the mean error of YouTube geotagging by half while retaining sixty percent coverage.}
\label{fig:rroc}
\end{figure}

We search $5,947,386,063$ tweets collected between 2014-02-01 and 2014-05-07 for (expanded) urls matching either \url{.*youtube.*} or \url{.*flickr.*}. We discard urls which were not shared by $3$ or more distinct users. This provided us with $447,441$ YouTube urls and $1,992$ Flickr urls. Using the public YouTube and Flickr APIs we were able to identify recording locations for $14,582$ videos and $284$ photos, respectively. The number of Flickr images is notably small. We suspect this is due to the fact that Twitter allows users to post images directly to \url{pic.twitter.com} which simplifies image sharing.

Our results on YouTube are promising. The error is substantially lower than on Flickr and is controllable via (\ref{eq:l1spread}), a point which is clarified in fig. \ref{fig:rroc}. Here, we show the response of the mean error to different restrictions on (\ref{eq:l1spread}) and can see that YouTube geotagging responds predictably to limitations on (\ref{eq:l1spread}). The fact that the YouTube API only provided location for $14,582$ of videos in our sample while the proposed method inferred locations for hundreds of thousands creates several new opportunities for geographic video retrieval.

\section{Conclusion}
\label{sec:conclusion}
In this work we have evaluated the effectiveness of geotagging online documents by examining only their url. The proposed method relies on locations of Twitter users who share links to the url in question and infers document content via robust estimates of the center and spread of user locations. We have evaluated our technique on multiple datasets built from text, photos, and videos.

Our results indicate that the accuracy of content-agnostic geotagging is, not surprisingly, heavily dependent on the content being geotagged. Our results indicate that social-network based geotagging is valuable for tweet content, YouTube videos, and multilingual news documents. Geotagging English-language news articles pertaining to worldwide events yields a large discrepancy against automated natural language processing systems. Propagating locations gleaned from Twitter to Tumblr and Flickr yielded questionable results. Additionally, we have found that the performance of our dispersion heuristic is varies with content and performs notably well on Twitter and YouTube. The ability to infer the geographic focus of other digital media, such as public Facebook posts, is a direction for future work. 


\section{Acknowledgments}
Supported by the Intelligence Advanced Research Projects Activity (IARPA) via Department of Interior National Business Center (DoI / NBC) contract number D12PC00285. The U.S. Government is authorized to reproduce and distribute reprints for Governmental purposes notwithstanding any copyright annotation thereon. The views and conclusions contained herein are those of the authors and should not be interpreted as necessarily representing the official policies or endorsements, either expressed or implied, of IARPA, DoI/NBE, or the U.S. Government.

%
\bibliographystyle{abbrv}
\bibliography{thebib}  

\begin{thebibliography}{10}

\bibitem{Ahern2007}
S.~Ahern, M.~Naaman, R.~Nair, and J.~H.-I. Yang.
\newblock World explorer: Visualizing aggregate data from unstructured text in
  geo-referenced collections.
\newblock In {\em Proceedings of JCDL '07}, pages 1--10, New York, NY, USA,
  2007. ACM.

\bibitem{Ahmed2013}
A.~Ahmed, L.~Hong, and A.~J. Smola.
\newblock Hierarchical geographical modeling of user locations from social
  media posts.
\newblock In {\em Proceedings of the 22Nd International Conference on World
  Wide Web}, WWW '13, pages 25--36, Republic and Canton of Geneva, Switzerland,
  2013. International World Wide Web Conferences Steering Committee.

\bibitem{Amitay2004}
E.~Amitay, N.~Har'El, R.~Sivan, and A.~Soffer.
\newblock Web-a-where: geotagging web content.
\newblock In {\em Proceedings of SIGIR '04}, pages 273--280. ACM, 2004.

\bibitem{bresson2013multiclass}
X.~Bresson, T.~Laurent, D.~Uminsky, and J.~H. von Brecht.
\newblock Multiclass total variation clustering.
\newblock In {\em Advances in Neural Information Processing Systems}, 2013.

\bibitem{compton2014geotagging}
R.~Compton, D.~Jurgens, and D.~Allen.
\newblock Geotagging one hundred million twitter accounts with total variation
  minimization.
\newblock {\em arXiv preprint arXiv:1404.7152}, 2014.

\bibitem{compton2013detecting}
R.~Compton, C.~Lee, T.-C. Lu, L.~De~Silva, and M.~Macy.
\newblock Detecting future social unrest in unprocessed twitter data:``emerging
  phenomena and big data''.
\newblock In {\em Intelligence and Security Informatics (ISI), 2013 IEEE
  International Conference on}, pages 56--60. IEEE, 2013.

\bibitem{Crandall2009}
D.~J. Crandall, L.~Backstrom, D.~Huttenlocher, and J.~Kleinberg.
\newblock Mapping the world's photos.
\newblock In {\em Proceedings of WWW '09}, pages 761--770, New York, NY, USA,
  2009. ACM.

\bibitem{Gelernter2013a}
J.~Gelernter and S.~Balaji.
\newblock {An algorithm for local geoparsing of microtext}.
\newblock {\em GeoInformatica}, 17(4):635--667, Jan. 2013.

\bibitem{Gelernter2013}
J.~Gelernter, G.~Ganesh, H.~Krishnakumar, and W.~Zhang.
\newblock Automatic gazetteer enrichment with user-geocoded data.
\newblock In {\em Proceedings of GEOCROWD '13}, pages 87--94, New York, NY,
  USA, 2013. ACM.

\bibitem{Goldstein}
T.~Goldstein and S.~Osher.
\newblock {The split bregman method for l1 regularized problems}.
\newblock pages 1--21.

\bibitem{Ireson2010}
N.~Ireson and F.~Ciravegna.
\newblock Toponym resolution in social media.
\newblock In {\em Proceedings of ISWC '10}, pages 370--385, Berlin, Heidelberg,
  2010. Springer-Verlag.

\bibitem{jurgens2013s}
D.~Jurgens.
\newblock That’s what friends are for: Inferring location in online social
  media platforms based on social relationships.
\newblock In {\em Seventh International AAAI Conference on Weblogs and Social
  Media}, 2013.

\bibitem{Kelm2013}
P.~Kelm, V.~Murdock, S.~Schmiedeke, S.~Schockaert, P.~Serdyukov, and
  O.~Van~Laere.
\newblock Georeferencing in social networks.
\newblock In {\em Social Media Retrieval}, pages 115--141. Springer, 2013.

\bibitem{leetaru2013gdelt}
K.~Leetaru and P.~A. Schrodt.
\newblock Gdelt: Global data on events, location, and tone, 1979--2012.
\newblock In {\em Paper presented at the ISA Annual Convention}, volume~2,
  page~4, 2013.

\bibitem{leetaru2013mapping}
K.~Leetaru, S.~Wang, G.~Cao, A.~Padmanabhan, and E.~Shook.
\newblock Mapping the global twitter heartbeat: The geography of twitter.
\newblock {\em First Monday}, 18(5), 2013.

\bibitem{Leidner2004}
J.~L. Leidner.
\newblock Toponym resolution in text: "{W}hich {S}heffield is it?".
\newblock In {\em Proceedings of SIGIR '04}, page 602, 2004.

\bibitem{Lieberman2010}
M.~D. Lieberman, H.~Samet, and J.~Sankaranarayanan.
\newblock Geotagging with local lexicons to build indexes for
  textually-specified spatial data.
\newblock In {\em Proceedings of ICDE '10}, pages 201--212, 2010.

\bibitem{mahmud2012tweet}
J.~Mahmud, J.~Nichols, and C.~Drews.
\newblock Where is this tweet from? inferring home locations of twitter users.
\newblock In {\em ICWSM}, 2012.

\bibitem{McCurley2001}
K.~S. McCurley.
\newblock Geospatial mapping and navigation of the web.
\newblock In {\em Proceedings of WWW '01}, pages 221--229, New York, NY, USA,
  2001. ACM.

\bibitem{OHare2013}
N.~O'Hare and V.~Murdock.
\newblock Modeling locations with social media.
\newblock {\em Inf. Retr.}, 16(1):30--62, Feb. 2013.

\bibitem{Overell2008}
S.~Overell and S.~R\"{u}ger.
\newblock Using co-occurrence models for placename disambiguation.
\newblock {\em Int. J. Geogr. Inf. Sci.}, 22(3):265--287, Jan. 2008.

\bibitem{ramakrishnan2014beating}
N.~Ramakrishnan, P.~Butler, S.~Muthiah, N.~Self, R.~Khandpur, P.~Saraf,
  W.~Wang, J.~Cadena, A.~Vullikanti, G.~Korkmaz, et~al.
\newblock 'beating the news' with embers: Forecasting civil unrest using open
  source indicators.
\newblock {\em arXiv preprint arXiv:1402.7035}, 2014.

\bibitem{Rattenbury2009}
T.~Rattenbury and M.~Naaman.
\newblock Methods for extracting place semantics from flickr tags.
\newblock {\em ACM Trans. Web}, 3(1):1:1--1:30, Jan. 2009.

\bibitem{Rudin1992}
L.~Rudin, S.~Osher, and E.~Fatemi.
\newblock {Nonlinear total variation based noise removal algorithms☆}.
\newblock {\em Physica D: Nonlinear Phenomena}, 60(1-4):259--268, Nov. 1992.

\bibitem{serdyukov2009placing}
P.~Serdyukov, V.~Murdock, and R.~Van~Zwol.
\newblock Placing flickr photos on a map.
\newblock In {\em Proceedings of the 32nd international ACM SIGIR conference on
  Research and development in information retrieval}, pages 484--491. ACM,
  2009.

\bibitem{Smith:2010:IDE:1840693.1928520}
L.~M. Smith, M.~S. Keegan, T.~Wittman, G.~O. Mohler, and A.~L. Bertozzi.
\newblock Improving density estimation by incorporating spatial information.
\newblock {\em EURASIP J. Adv. Signal Process}, 2010:7:1--7:12, Feb. 2010.

\bibitem{VanLaere2014}
O.~Van~Laere, J.~Quinn, S.~Schockaert, and B.~Dhoedt.
\newblock Spatially aware term selection for geotagging.
\newblock {\em IEEE Trans. on Knowl. and Data Eng.}, 26(1):221--234, Jan. 2014.

\bibitem{VanLaere2011}
O.~Van~Laere, S.~Schockaert, and B.~Dhoedt.
\newblock Finding locations of flickr resources using language models and
  similarity search.
\newblock In {\em Proceedings of ICMR '11}, pages 48:1--48:8, New York, NY,
  USA, 2011. ACM.

\bibitem{Vardi2000}
Y.~Vardi and C.~H. Zhang.
\newblock {The multivariate L1-median and associated data depth.}
\newblock {\em Proceedings of the National Academy of Sciences of the United
  States of America}, 97(4):1423--6, Feb. 2000.

\bibitem{Xu_websci2014}
J.~Xu, T.-C. Lu, R.~Compton, and D.~Allen.
\newblock Quantifying cross-platform engagement through large-scale user
  alignment.
\newblock In {\em ACM Web Science Conference}, 2014.

\bibitem{Yamaguchi:2013:LUL:2512938.2512941}
Y.~Yamaguchi, T.~Amagasa, and H.~Kitagawa.
\newblock Landmark-based user location inference in social media.
\newblock In {\em Proceedings of the First ACM Conference on Online Social
  Networks}, COSN '13, pages 223--234, New York, NY, USA, 2013. ACM.

\bibitem{Yin2011}
Z.~Yin, L.~Cao, J.~Han, C.~Zhai, and T.~Huang.
\newblock {Geographical topic discovery and comparison}.
\newblock In {\em Proceedings of WWW '11}, page 247, New York, New York, USA,
  2011. ACM Press.

\bibitem{zhu2002learning}
X.~Zhu and Z.~Ghahramani.
\newblock Learning from labeled and unlabeled data with label propagation.
\newblock Technical report, Technical Report CMU-CALD-02-107, Carnegie Mellon
  University, 2002.

\end{thebibliography}
%
%

\end{document}